# Determining the thickness and modulus of the nano-metric interphase region using macroscopic modulus data in exfoliated polymer-clay nanocomposites

H. Goodarzi Hosseinabadi [a,*], Kh. Khederlou [b], J. Payandeh Peyman [c], R. Bagheri [a]

[a] Polymeric Materials Research Group, Department of Materials Science and Engineering, Sharif University of Technology, P.O.Box 11155-9466, Tehran, Iran

[b] Institute for Nanoscience and Nanotechnology, Sharif University of Technology, Tehran, Iran

[c] Department of Mechanical Engineering, Bu-Ali Sina University, Hamedan, Iran

* Corresponding Author. Email address: hosseina@mit.edu ; hgoodarzy@gmail.com

## Abstract

This paper is an attempt to predict the thickness and modulus at the nanometric interphase region with the knowledge on the macroscopic Young's modulus in the polymer-clay nanocomposites. First, a systematic design of 20'160 linear elastic finite element simulations are used to derive an analytical equation between the interphase thickness, the interphase modulus, the nanoclay content and the macroscopic Young's modulus of the nanocomposite. Four calibration parameters in this equation are calculated based on the reported data about NYLON6-MMT nanocomposite. Next, an analytical nanoscale equation is developed to satisfy the necessary boundary conditions at the nanometric interphase. Finally, the interphase thickness and modulus are calculated by the intersection between the two equations. The validity of predictions are examined using two distant data sets in literature. The predicted interphase thicknesses are consistent with the experimental reports which situate around 2-8 nm in this nanocomposite. The presented approach can be used in design and prediction of mechanical properties in wide range of isotropic triple-phasic systems with structural features similar to polymer clay nanocomposites.

# 1. Introduction

Where roughly 95% of plastic production belongs to the multi-phase systems –including blends, composites and foams, it seems that polymer clay nanocomposites (PCNs) are outpacing their rivals. This is mainly because in recent decades the dispersion of nanoclay in the forms of exfoliated or intercalated in polymer matrixes has emerged many applications and research interests [1–5]. In this ground, the plate like montmorillonite (MMT) is known as one of the most popular choices among broad class of nanoclays. When dispersed in a polymer environment, it dramatically improves the properties of PCN even at a few weight percent [6–8]. At molecular level, the surfactant interactions at the interface between the nanoclay and polymer matrix suppresses the polymer chains mobility resulting in formation of an interphase characterized by its thickness and elastic modulus –the parameters that describe the PCN's mechanical response [9]. If the interphase modulus and thickness can be precisely approximated, these information can be used to design materials with tailored mechanical properties made of such a three-phase system including the matrix, the nanoclay and the interphase. Despite its importance and several thousands of citation to the work by Kojima et al [10] who firstly introduced the concept of interphase in NYLON6-MMT system, the relationship between the interphase thickness and modulus is still not clear. Among the studies on mechanical behavior of these nanocomposites, limited number have directly attended to the effect of interphase on the PCN mechanical response [9–14]. This is not surprising since remarkable limitations have been associated with experimental measurements of the interphase modulus and thickness [9]. In this ground, computational solutions such as finite element (FE) [15, 16] obtain strategic importance. So far, several research groups have used the FE to predict the elastic properties of NYLON6_MMT nanocomposite by conducting either 2D [14, 17], or 3D simulations [2, 18]. Nevertheless, the correlation of interphase modulus variations to the interphase thickness variations is remained under debate [19]. In this paper we extend a previous work [20] with conducting a larger number of simulations to build a structure-property relationship for describing the interphase thickness and modulus relation using the knowledge on material macroscopic modulus. To this aim, a total of 20'160 statistically precise simulations are conducted to formulate the structural influence of the interphase properties and nanoclay loading on the macroscopic modulus of the nanocomposite. Similar approach can be used for the large class of three phase systems with similar geometry and aspect ratio for the geometries of the constituents.

## 2. Method

First of all, large number of FE simulations were designed. The numerical results were used to develop an analytical $E_i$-$t_i$ equation between the investigated parameters. Then, a second $E_i$-$t_i$ equation was developed to satisfy necessary boundary conditions at nanometric scale. Eventually, the interphase thickness and modulus calculated by analyzing the equations intersection.

*2- 1- Designing FE simulations*

Representative Volume Element (RVE) is a representative volume of the material that has statistically the same mechanical response as the bulk sample [21–23]. In order to statistically make sure that the simulation results have a reasonable accuracy, the minimum number of required random patterns to obtain a 95% confidence level, 0.05 tolerances and 0.125 deviations (estimated value) was considered as large as 24 patterns in this research. More statistical details about this method is presented elsewhere [21]. Accordingly, an algorithm written in MATLAB™ was utilized to generate 24 random RVEs for each configuration. Each configuration was comprised of 4 different nanoclay weight percents, 7 different interphase thicknesses and a variety of 15 different interphase modulus. The range of these variables were accustomed to encompass the range of data reported in literature [7]. Figure 1 shows the design of simulations. Each RVE was loaded in both horizontal and vertical directions to examine the isotropy of volume elements.

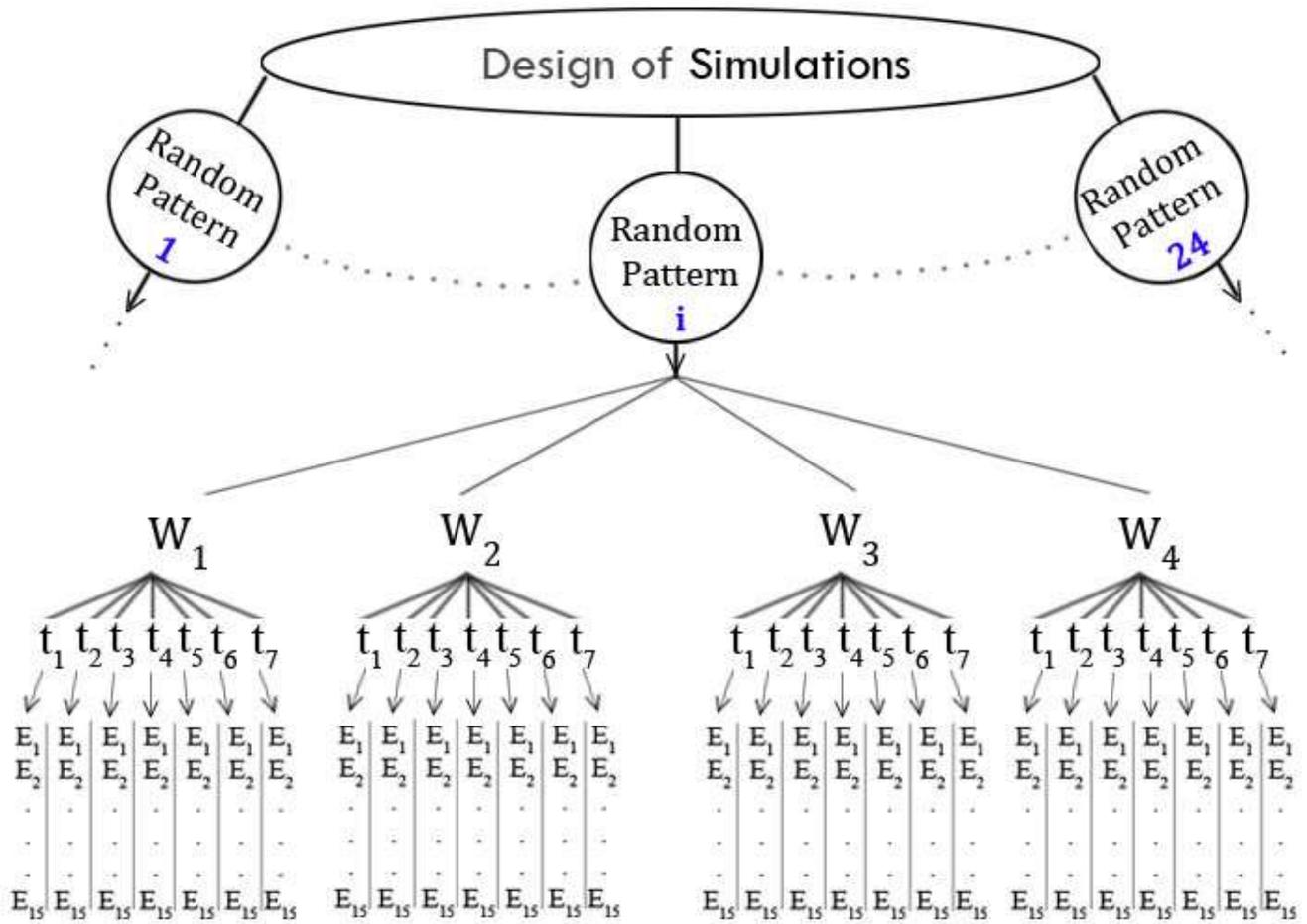

**Figure 1- Design of simulations**: range of variables that customized to include the available data in literature **Wt**[%]=[1.6, 3.2, 4.5, 7.2], **ti**[nm]=[1, 2, 3, 4, 5, 6, 7], **Ei** [GPa]=[0.7, 1.4, 2.1, 2.8, 5.5, 8.3, 11, 13.8, 16.5, 19.3, 22, 24.8, 27.5, 30.3, 33].

Figure 2 illustrates the boundary conditions imposed for typical horizontal and vertical loadings, comparable to the kinematic uniform boundary condition (KUBC) which in similar cases is employed [22].

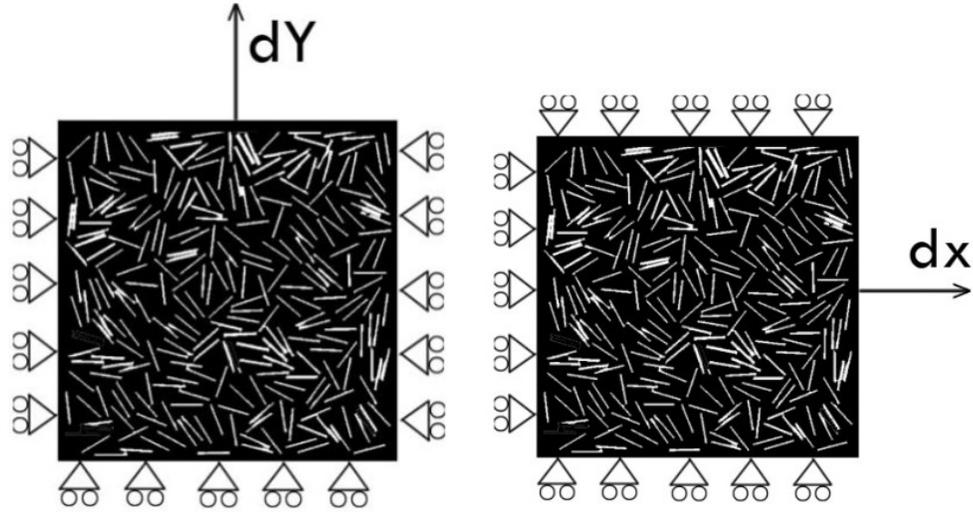

**Figure 2- Boundary Conditions**: for horizontal and vertical loadings.

In horizontal loading, the Hook's elasticity theory for plane stress conditions together with the mesh type of PLANE182 in Ansys$^{TM}$ library was used to calculate the material Young's modulus according to Equation 1.

$$E_{xx} = \frac{\sigma_x}{\varepsilon_x}(1-v_{xy}^2) \tag{1}$$

In this equation, $\sigma_x$ calculated by dividing the reaction force summation to the area perpendicular to the force direction, and $\varepsilon_x$ -the infinitesimal elastic strain in X direction, calculated by:

$$\varepsilon_x = \frac{dx}{L_x} \tag{2}$$

where $L_x$ is the RVE length in loading direction. In addition, to evaluate the isotropy in nanoclay distribution the RVE's elastic response calculated in Y direction by Equation 3.

$$E_{yy} = \frac{\sigma_y}{\varepsilon_y}(1-v_{yx}^2) \tag{3}$$

where $\sigma_y$ calculated by dividing the reaction force summation to the area perpendicular to the force direction, and $\varepsilon_y$ -the infinitesimal strain in Y direction, calculated by:

$$\varepsilon_y = \frac{dy}{L_y} \tag{4}$$

Here $L_y$ denotes the RVE length in loading direction and dy is the deformation in corresponding direction. Finally, the average of the Young's modulus that were obtained in both loading conditions was considered as the RVE's modulus. In this way, a total of 20'160 simulations were performed to precisely evaluate the influence of variables (the nanoclay content and the interphase properties) on the RVE Young's modulus.

According to previous reports the results obtained by a $1 \times 1 \mu m^2$ size RVE converges to the results obtained by larger RVE sizes [24, 25]. In this work, we used a similar RVE size and the mesh size was adjusted based on a mesh size sensitivity analysis that was separately performed for the most critical configuration with highest clay content and thinnest interphase thickness. Additionally, it was assumed that the interphase lay on the top and bottom faces of the nanoclay to form a three-layer stack (Figure 3).

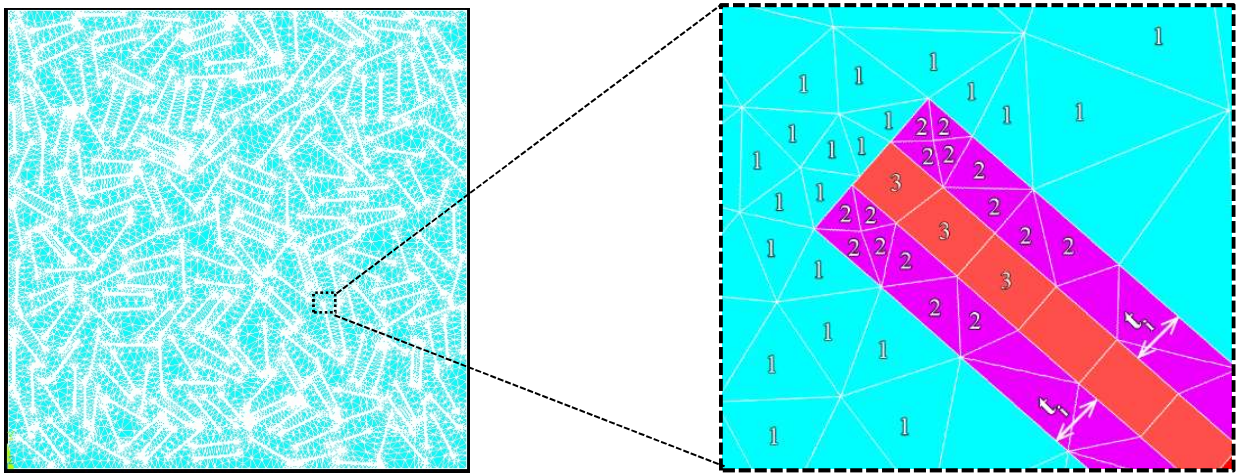

**Figure 3- Meshed volume element:** The meshed RVE and the magnified 3-layer stack. '1' denotes the polymer matrix elements, '2' denotes the interphase elements and '3' denotes the nanoclay layer. (Coarse mesh size is shown at here to improve the quality of figure).

The material parameters were considered according to Table 1. Note that the silicate length and modulus is not well defined in literature and are approximated from literature [26, 27].

Table 1- Used material parameters as the input in simulations.

| Material | E (GPa) | Poisson Ratio | Density (g/cm³) | Assumed MMT Length/Thickness |
|---|---|---|---|---|
| Nano-clay (MMT) | 178 | 0.28 | 2.35 | 100/1 |
| NYLON6 (HMW) | 2.75 | 0.35 | 1.14 | - |

*2- 2- Deriving macro-scale equation*

According to the results obtained by the simulations Equation 5 was derived to describe the relation of nanoclay content, interphase modulus and interphase thickness to the macroscopic modulus.

$$E_C^* = \Phi \left\{ A \sinh(\gamma t_{in})^M \ln(E_{in}) + B E_{clay}^* \right\} \tag{5}$$

where A, B, γ and M are dimensionless calibration parameters that would be optimized for providing the best-fit to the FE results. These parameters are only related to the input variables that summarized in Table 1 and are constant for each PCN system. The $\Phi$, $t_{in}$ and $E_{in}$ variables demonstrate the nanoclay volume percent, the normalized interphase thickness and the normalized interphase modulus, respectively. The $t_{in}$, the $E_{in}$, the normalized difference of the nanocomposite Young's modulus ($E_C^*$) and the normalized difference of the nanoclay Young's modulus ($E_{clay}^*$) are defined by Equations 6.

$$E_C^* = \frac{E_C - E_m}{E_m} \; ; \; E_{clay}^* = \frac{E_{clay} - E_m}{E_m} ; \; t_{in} = \frac{t_i}{t_{clay}} \; ; \; E_{in} = \frac{E_i}{E_m} \tag{6}$$

where $E_C$ is the nanocomposite macroscopic elastic modulus, $E_m$ denotes the matrix polymer modulus and $E_{clay}$ demonstrates the nano-clay modulus. Accordingly, the $E_i$-$t_i$ relationship can be furnished as:

$$E_i(t_i) = E_m \exp\left(\frac{P_{EN} - BE^*_{clay}}{A \sinh(\gamma \frac{t_i}{t_{clay}})^M}\right) \text{, where } P_{EN} = \frac{E^*_C}{\Phi} \quad (7)$$

This equation supports all of the data generated by several thousands of simulations in previous step which provided best-fit using the least square error technique. The ordinary least square method is a standard approach in regression analysis with approved application in data fitting. The "best-fit" in the least-squares sense minimizes the sum of squared residuals, where a residual is defined here as the difference between an observed value from simulation results and the calculated value by the analytical equation. The form and type of the used functions in this equation were approximated by analyzing the simulation outputs and were investigated among a large number of equation types in Matlab™ and Origin™ libraries. In our explorations, the presented form in Equation 7 was the only combination that minimized the least square error residuals with reasonable number of calibration parameters.

Note that in this equation "$P_{EN}$" is the parameter of efficiency of nano-clay. Whenever $P_{EN}$ opts a higher value a higher nanocomposite modulus is obtained per nano-clay volume percent, which can be addressed as the higher efficiency of the clay. It is worthwhile to note that in each PCN system the $P_{EN}$ value can be changed with increasing the applied nanoclay content. Depending on the processing conditions, the efficiency of added nanoclay ($P_{EN}$ value) may get increased or may get decreased by increasing the clay content. In our model, after determination of the calibration constants for each PCN system, the experimental value for the $P_{EN}$ parameter should be calculated from the results of the tensile test measurements by dividing the $E^*_C$ (computed by replacing the measured Young's modulus into the Equation 6) on the consumed nanoclay volume percent. In this work we have used two separate date sets as are reproduced in Figure 4.

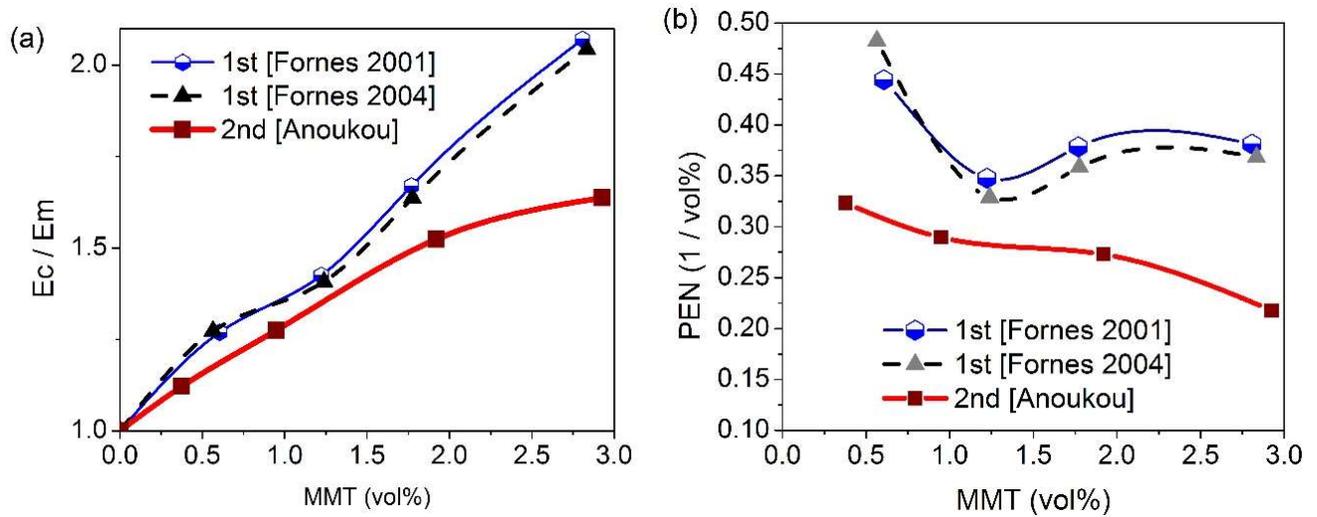

**Figure 4- Comparison of two experimental data reported on NYLON6_MMT system:** variations of a) the normalized PCN modulus & b) the parameter of efficiency of nanoclay, versus the nanoclay volume percent.

These data sets are carefully selected as their reported material properties are comparable to the material parameters that were used in our simulations according to Table 1 and there exist several experimental reports to validate our model. The first set, includes two reported data set [24, 27] for the macroscopic modulus of a melt compounded NYLON6_MMT mixture which the nanoclay surfaces were similarly functionalized and processed by same equipment. Each data in this figure represents an average of the results for 6 tensile tests. The standard deviations were typically of the order of 0.04 for the measured Young's modulus. The normalized nanocomposite modulus is enhanced more or less linearly with increasing the nanoclay content. Proportionally, the calculated parameter of efficiency of nanoclay ($P_{EN}$ as defined by Equation 7) remains almost constant with increasing the clay content except at the very low clay loadings where agglomeration or overlapping of neighboring stress fields is minimal and thus, the highest clay efficiency is anticipated.

For the 2$^{nd}$ set, an average data of 5 tensile tests on the same PCN system is reproduced here from the results provided by a separate compounding system [21]. These data has experienced different processing conditions in comparison with the 1st set. The PEN parameter in this case shows a meaningful decay by increasing the clay content (Figure 4-b). The unavoidable decay at high clay concentrations is reported in other researches [28]. The low effectiveness at high clay contents is attributed to higher probability of clay agglomeration in previous works however our model relates this effect to the structural effects mainly caused by the properties of the interphase region.

As soon as this experimental $P_{EN}$ values are replaced in Equation 7, the interphase thickness and modulus relationship is defined explicitly in each nanoclay volume percent. Remind that calibration parameters where previously calculated for this specific PCN system by the bet fitting algorithm. As Equation 7 only uses the macroscopic modulus of the nanocomposite to describe the $E_i$-$t_i$ relation herein we address it as the macro-scale equation.

*2- 3- Deriving nano-scale equation*

At nanoscopic scale the following boundary conditions were supposed to get satisfied by the interphase thickness and modulus:

$$\begin{cases} t_i^{local} = 0 \rightarrow E_i^{local} = E_{clay} \\ t_i^{local} = t_i \rightarrow E_i^{local} = E_m \end{cases} \quad (8)$$

where $t_i^{local}$ and $E_i^{local}$ denote the distance from the nanoclay surface ($t_i^{local} \leq t_i$) and the respective interphase modulus. It was supposed that a coherent interface exists between the nanoclay and the interphase region. Accordingly, beside the clay surface the interphase rigidity should locally converge to that of the clay. On the other side, the interphase modulus should converge to that of the matrix polymer. It was also considered that Equation 9 should be satisfied by the average modulus of the interphase region.

$$\begin{cases} E_i\big|_{t_i \to 0} = \dfrac{(E_m + E_{clay})}{2} \\ E_i\big|_{t_i \to \infty} = E_m \end{cases} \text{ where } E_i = Average(E_i^{local}) = \dfrac{\int_0^{ti} E_i^{local} dt_i^{local}}{t_i - 0} \tag{9}$$

The first condition means that the average interphase modulus can be approximated by a linear correlation at small interphase thicknesses. The later condition means that the average of the interphase modulus will converge to that of the polymer at very thick interphase layers. Eventually, Equation 10 demonstrates the final nanometric relation that satisfies all above boundary conditions.

$$E_i(t_i) = E_{clay} + (E_m - E_{clay})\left(\dfrac{1 - t_i - e^{-t_i}}{t_i(e^{-t_i} - 1)}\right) \tag{10}$$

*2- 4- Prediction of the actual $E_i$ and $t_i$*

The actual modulus and thickness of the interphase region should satisfy both macro-scale and nano-scale equations. To this aim the intersection of both equations is extracted by plotting Equation 7 & 10.

## 3. Results and discussion

Owing to the parametric design language in the Ansys™, the running of FE simulations for each of the 24 configuration shown in Figure 1 finalized in dozen hours. Figure 5 shows several minimum, maximum and average Young's modulus data that were obtained from 24 randomly created pattern which all include a 1 nm interphase thickness and 0.5 and 2.5 nanoclay volume percent.

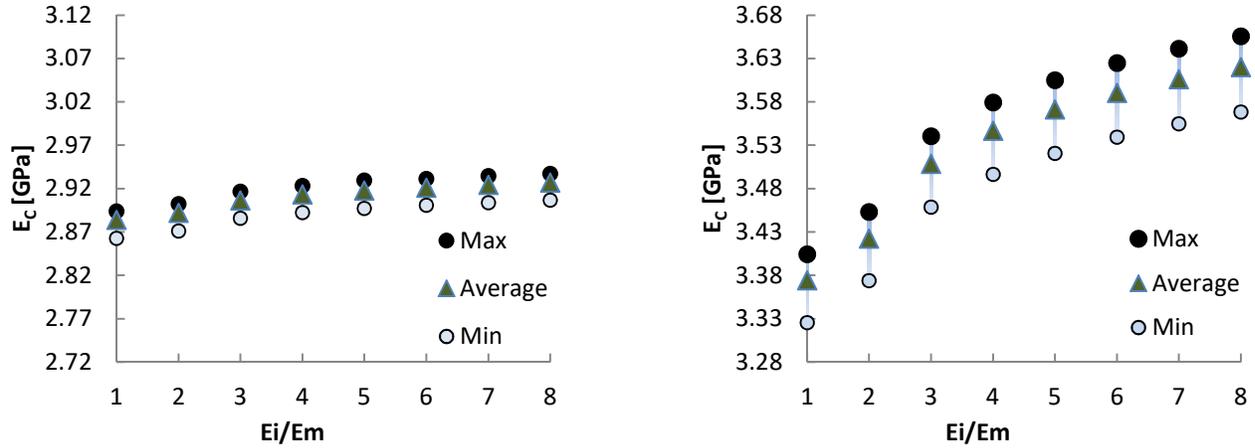

**Figure 5- Statistical evaluation:** scattering of the simulated nanocomposite modulus among 24 random RVEs versus several interphase modulus; the interphase thickness is set 1nm and the clay content is set: a) 0.5, b) 2.5 vol %.

It is clear that increasing the nanoclay volume percent will enhance the scattering of simulated nanocomposite modulus, however, the range of scatterings will still remain small and the results show an acceptable reproducibility. The standard deviation of the simulated Young's modulus among all of the 24 simulated patterns was well below 0.1. The reasonably small standard deviation of the modulus obtained from the 24 random sample proves the acceptable quality of the written algorithm in MATLAB™ to generate fully isotropic RVEs which was required for simulation of the exfoliated PCN microstructure. As the standard deviation of calculated Young's modulus in horizontal and vertical loadings from their average was situated within [0.006, 0.02] interval (well below our first approximation of 0.125 value at the design of experiments stage) it can be inferred that well below 24 random RVEs would be enough to achieve a statistically reliable response in these elastic simulations [20]. The calibration constants of Equation 7 that obtained by the Least Square technique are summarized in Table 2.

Table 2- Best fitted calibration parameters in Equation 7.

| *Parameter* | $A$ | $BE^*_{cly}$ | $\gamma$ | $M$ |
|---|---|---|---|---|
| Value | 0.076 | 0.088 | 0.206 | 0.838 |

Figure 6 compares the analytical macro-scale equation with the numerical simulation results to show the accuracy of the best-fitting procedure. Red points describe the numerical results and blue lines represent the predictions of the analytical equation. As can be seen, the analytical macro-scale equation (Equation 7) has good accordance with the simulation results.

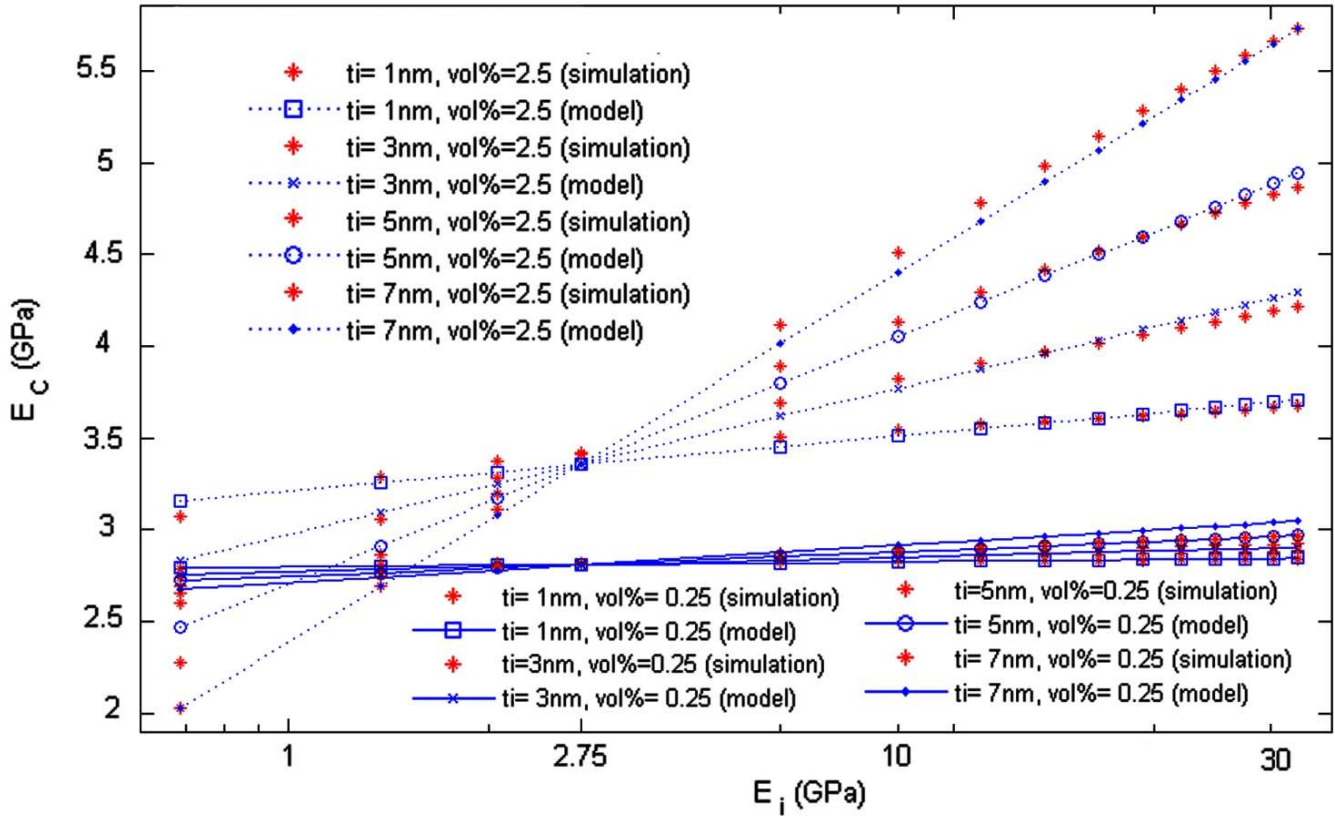

**Figure 6- Comparison of the best fitted model with the simulation results:** Nanocomposite modulus versus the interphase modulus for a variety of interphase thicknesses and two different clay contents. ('+' symbols describe the simulation results and continuous lines represent the Macro-scale equation response)

Figure 7 magnifies the influence of the clay content on the predictions of the model and original simulation results.

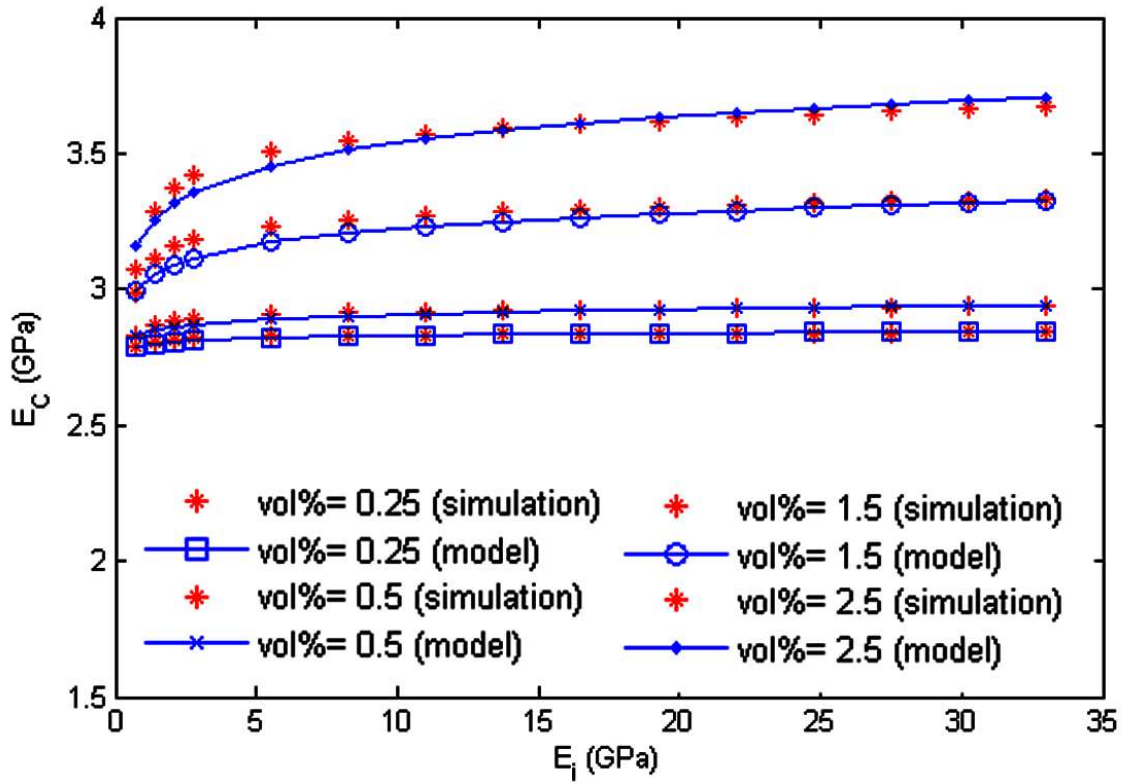

**Fig 7- Comparison of the best fitted model with the simulation results:** Nanocomposite modulus versus the interphase modulus only for $t_i$= 1nm and a variety of clay contents. ('+' denotes the simulation results and continuous lines represent the value of the macro-scale equation)

As shown, increasing the nanoclay content not only increases the macroscopic modulus but also magnifies the influence of interphase modulus on the nanocomposite modulus. In both cases, a logarithmic relationship between Ei and nanocomposite modulus is observed. Note that in Figure 6, the graphs met each other at Ei=E$_{matrix}$ since the interphase had similar properties of the matrix in a constant nanoclay content but at here when Ei=E$_{matrix}$ the graphs do not intersect since different nanoclay contents provide different strengthening effects. This difference can be predicted by the rule of mixtures.

The behavior of the nanoscale equation is depicted at Figure 8 where the nano scale equation -that is related to the average properties of the interphase, together with the local interphase equation –that is related to the local properties of the interphase are illustrated.

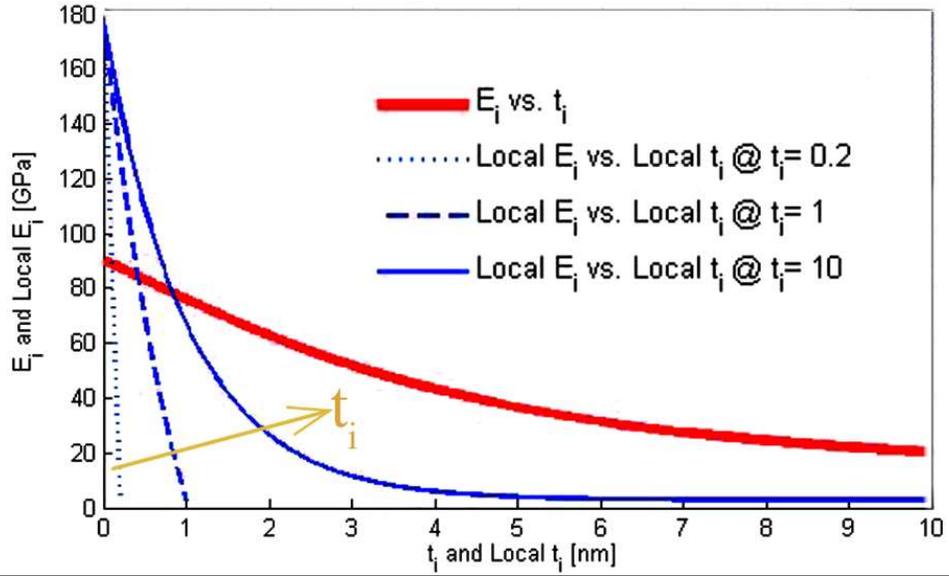

**Figure 8- Nano scale equation:** illustration of the local and averaged interphase modulus versus the interphase thickness relationship.

In this graph, the dashed lines depict the $E_i^{local} - t_i^{local}$ relationship for a variety of interphase thicknesses according to Equation 10. The thick continuous curve shows the $E_i$-$t_i$ relationship based on Equation 10. The former satisfies the boundary conditions declared by Equation 8 and the latter satisfies the boundary conditions stipulated by Equation 9. The dashed lines include local information about the interphase, while the continuous line is a volume average of the local variations thus it is a more stable measure. As the interphase thickness increases, the average interphase modulus decreases. This behavior is intentionally included in formulation due to a sense obtained from the thermodynamics of the interphase formation.

To achieve the intersection point of macro- and nano- scale equations (denoted by $E_i^*$ and $t_i^*$), the parameter of efficiency of nanoclay (PEN) is extracted according to the 1$^{st}$ and 2$^{nd}$ data in Figure 4 . When replaced in Equation 7, the macroscale equation together with the nanoscale equation can be plotted.

Figure 9 shows the range of predictions for the interphase thickness & modulus.

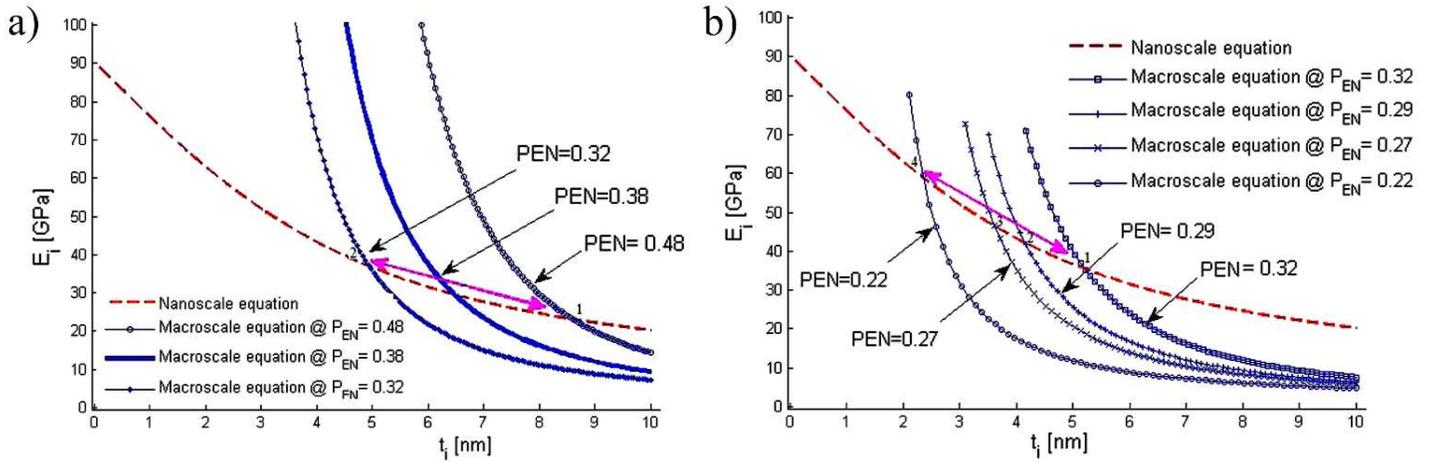

**Figure 9- variations of $E_i$ versus $t_i$ based on the nanoscale equation (discontinuous line) and the macroscale equation (continuous lines at different $P_{EN}$ values)**: using a) the 1st [15,19] & b) the 2nd [21] experimental data set. The double side arrows represent the predicted range of actual $E_i$ and $t_i$.

It is obvious that the interphase thickness and the interphase modulus obtained by the 1st data set lay within 6.5±2 nm and 29±7 GPa intervals, respectively. It seems that in this case -where the efficiency of nanoclay do not show a meaningful dependency to the nanoclay percent, one may prefer to report an average of $t_i^*$ and $E_i^*$ independent of the nanoclay content.

For the 2nd data set, decreasing the value of effectiveness of clay ($P_{EN}$) will proportionally shift the intersection point to a lower interphase thicknesses and higher interphase modulus. Figure 10 shows the relationship between the interphase thickness/modulus to the $P_{EN}$ values and the nanoclay contents.

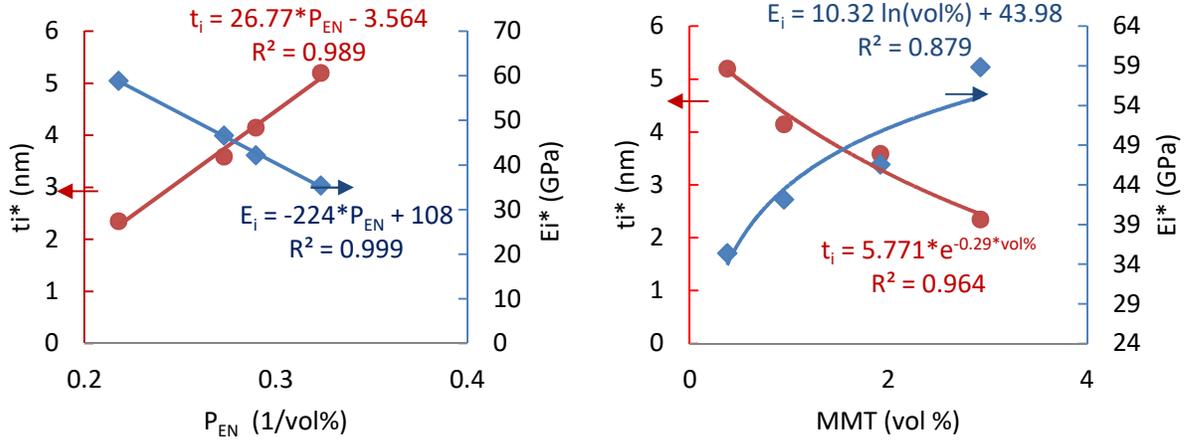

**Figure 10- Dependency of the predicted interphase thickness and modulus**: on the (a) $P_{EN}$ parameter and (b) clay content.

Accordingly, the interphase thickness decreases from around 5.2 nm to 2.5 nm by increasing the nanoclay volume percentile. A linear accordance between $t_i^*$ & $P_{EN}$ and between $E_i^*$ & $P_{EN}$ value is observed, which could not be expected by reviewing the complex form of only macroscale or only nanoscale equations. It may underscore the importance of $P_{EN}$ parameter to formulate the elastic mechanical responses in PCN systems. When converted to nanoclay volume percent, a non-linear function best-fits to the predicted $E_i$ and $t_i$ data as are merged within the Figure 10.

Note that the analytical equations illustrated on Figure 10 are obtained for the 2$^{nd}$ data set, while the 1$^{st}$ data set could not predict an analytical $t_i^* - E_i^*$ relation but provides an approximated range. It seems that the $t_i^* - E_i^*$ relationship is significantly dependent on the implemented processing technique.

Unfortunately, the available experimental reports about the interphase properties are not large enough to validate all of the predicted interphase thickness and modulus at here. Therefore, we summarize a combination of predicted data by dynamic mechanical analysis, nano indentation tests and molecular dynamics predictions. Mesbah et al. [29] have employed differential scanning calorimetry and dynamic mechanical analysis then have estimated the interphase thickness ($t_i^*$) around 2-3 nm. Xu et al. [30] have employed molecular dynamics technique and have predicted the interphase thickness as large as 3 nm. Sikdar et al. [14] have utilized atomic force microscopy and nano-indentation tests to estimate the $t_i^*$ as large as 2.5 nm and the interphase thickness ($E_i^*$) around 14 GPa. According to Shelly et al.[9] and the loss and storage modulus data, the $t_i^*$ can be predicted as large as 8.5 nm and $E_i^*$ around 28 GPa [18].

In the current work for the 1st set of processed NYLON6_MMT the $t_i^*$ is estimated 6.5±2 nm and the $E_i^*$ is predicted 29±7 GPa. For the 2nd data set, the $t_i^*$ - $E_i^*$ relation is calculated analytically in the range of 2.5< $t_i^*$ <5.2 nm and 35< $E_i^*$ <55 GPa. As can be inferred, the interphase thicknesses computed by our approach is comparable to the order of reported results. The predicted interphase modulus however is somehow out of the reported experimental range. Note that we have computed the average elastic modulus at the interphase region (according to equation 9) that is different from the local elastic modulus commonly measured in experiments. Most importantly, all experimental approaches measure the rigidity of an interphase at a truncated surface. To the best of our knowledge, the presence of dangling bonds on the truncated surficial layer can alleviate the stiffness of material at the surface in comparison with the actual interphase modulus. This effect can lead to an underestimation of the actual interphase modulus that is measured for example by the nano indentation and the model predictions on the interphase properties remain more or less logical. It is also worth noticing that taking into account 3-Dimentional RVEs may result in more realistic results, however, not only the computational costs will increase significantly but also the availability of a handy macro-scale equation (Equation 7) may not be possible. For convenience, the required steps to predict the interphase thickness and modulus by the presented approach are summarized at below:

1- Determining the calibration parameters in macro-scale equation:

a- Conducting a number of FE simulations in order to set the A, B, γ and M calibration parameters. [These parameters were related to the constituents of the PCN system (see Table 1) and were derived with assumption of randomness in nanoclay distribution. In case of NYLON6_MMT system these parameters were reported in Table 2.]

b- Conducting tensile test to calculate the experimental modulus of PCN at a known clay content. This data was then used to compute the efficiency of nanoclay parameter by Equation 6 and to make the macro scale equation explicitly determined.

2- Deriving the nano-scale equation that satisfied a series of boundary conditions.

3- Determining the intersection of macro- and nano-scale equations in order to find the $t_i^*$ - $E_i^*$ couple as the characteristic feature of the interphase.

The presented approach can be used for designing new polymer clay nanocomposites with consideration of the influence of the processing stage. For instance, the dependency of the interphase elastic modulus and the interphase thickness to the nanoclay loading (as same as the relationships that merged in Figure 10) can be obtained in different PCN systems. These dependencies can be imported in a finite element package to explore the mechanical performance of complex structures that can be made of such a three-phase system including the matrix, the nanoclay and the interphase material.

## 4. Conclusions

In this research, an approach was presented to calculate the nanoscopic characteristics of interphase region in a fully exfoliated polymer-clay nanocomposite. First, the relationship between the thickness and modulus of the interphase was formulated. Statistically precise simulations were conducted to calibrate the coefficients of a new analytical equation that formulated the interphase properties using the PCN macroscopic modulus. Then a nano-scale equation was derived and coupled with previous equation. Finally, the interphase thickness and modulus obtained through the intersection of equations. The predictions of the presented approach for the interphase thickness laid within the measured values that were provided by DMA and nano indentation experiments. The predicted interphase modulus however overestimated the experimentally reported data that may be related to the error associated with experimental measurements. This approach can be used for designing and prediction of properties in plolymer-clay nanocomposites. Moreover, the relationship between elastic modulus and thickness of the interphase can be used in a finite element package for exploration of the mechanical behavior in complex structures that can be made of this PCN using the same processing equipment.